\documentclass[%
 reprint,
 amsmath,amssymb,
 aps,
]{revtex4-1}

\usepackage{graphicx}
\usepackage{dcolumn}
\usepackage{bm}
\usepackage{subfigure}
\usepackage[english]{babel}

\begin{document}

\preprint{APS/123-QED}

\title{Spectral evolution with doping of an antiferromagnetic Mott state}

\author{Wu, Huan-Kuang$^{1,2}$ and Lee, Ting-Kuo$^1$}
\affiliation{%
$^1$Institute of Physics, Academia Sinica, Nankang Taipei 11529, Taiwan\\
$^2$Department of Physics, National Taiwan University, Daan Taipei 10617, Taiwan
}%

\date{\today}

\begin{abstract}
Since the discovery of half-filled cuprate to be a Mott insulator, the excitation spectra above the chemical potential for the unoccupied states has attracted many research attentions.  There were many theoretical works using different numerical techniques to study this problem, but many have reached different conclusions. One of the reasons is the lack of very detailed high-resolution experimental results for the theories to be compared with.   Recently, the scanning tunneling spectroscopy (STS)\cite{cai2015visualizing,ye2013visualizing} on lightly doped Mott insulator with an antiferromagnetic (AFM) order found the presence of in-gap states with  energy of order half an eV above the chemical potential. The measured spectral properties with doping are not quite consistent with earlier theoretical works.  In this paper we perform a diagonalization method on top of the variational Monte Carlo (VMC) calculation to study the evolution of AFM Mott state with doped hole concentration in the Hubbard model (HM). Our results found in-gap states that behave similarly with ones reported by STS. These in-gap states acquire a substantial amount of dynamical spectral weight transferred from the upper Hubbard band. The in-gap states move toward chemical potential with increasing spectral weight as doping increases. Our result also provides information about the energy scale of these in-gap states in relation with  the coulomb coupling strength U.

\end{abstract}

\pacs{Valid PACS appear here}
\maketitle


\section{\label{sec:level1}Introduction}

The spectral properties of the Mott insulators as a function of doping has been one of the key issues in studying the physics of high Tc cuprate superconductors. There are many theoretical works on this topic include exact diagonalization (ED)\cite{PhysRevB.45.10032,PhysRevB.46.3183,PhysRevB.48.3916}, quantum Monte Carlo (QMC)\cite{PhysRevLett.72.705} method, dynamical cluster approximation (DCA)\cite{0295-5075-56-4-563}, dynamical mean field theory (DMFT)\cite{PhysRevLett.77.131,PhysRevLett.80.2393,PhysRevB.73.165114,PhysRevB.81.235133,PhysRevB.89.155134}, and also real-space Green's function approach\cite{PhysRevB.83.115117}. According to their results, generally, a clear spectral weight transferred from the upper Hubbard band (UHB) to lower Hubbard band (LHB) that situated at the chemical potential can be seen as doping increases. But the details are different. In Ref. \cite{PhysRevB.46.3183,PhysRevLett.77.131,0295-5075-56-4-563,PhysRevLett.80.2393,PhysRevB.73.165114,PhysRevB.81.235133}, there are in-gap signals found. These signals become farther away from chemical potential together with the UHB as doping increases. There were also several results reported by using the cluster perturbation theory (CPT).  For the electron-doped case\cite{PhysRevB.90.245102}, an in-gap state was found at the bottom of UHB, which is in agreement with the in-gap states seen by STS on Ca$_2$CuO$_2$Cl$_2$\cite{ye2013visualizing} near an impurity. For the hole-doped case\cite{PhysRevLett.92.126401,PhysRevLett.94.156404,PhysRevB.74.235117,PhysRevLett.108.076401,krinitsyn2014cluster}, there were low-energy in-gap states with energy less than $0.2U$. On the experimental side, the x-ray absorption spectra (XAS) \cite{PeetsPhysRevLett.103.087402,ChenPhysRevB.88.134525} have observed the spectral weight transferred from UHB to LHB as doping increases but the broadness of the peaks makes it difficult to make detail comparison.

Recently, STS\cite{cai2015visualizing,ye2013visualizing}  reported for  very underdoped cuprates with a long range AFM order probed the spectral function across the charge-transfer (CT) gap. It found some  new results unexpected from earlier theoretical works.  In Ref. \cite{cai2015visualizing}, finite density of states appear inside and throughout the CT gap and only a small energy range at the chemical potential remains empty of spectral weight after holes are doped into the sample. The  peak positions of these in-gap states at lower doping can be above chemical potential up to $40\%$ of CT gap.  This is much larger than the in-gap states reported earlier\cite{PhysRevLett.92.126401,PhysRevLett.94.156404,PhysRevB.74.235117,PhysRevLett.108.076401,krinitsyn2014cluster,PhysRevB.90.245102} for HM calculations with U representing the CT gap. Besides the presence of in-gap states, when the system is doped with holes, there is also a systematic evolution in the spectral weight distribution or local density of states (LDOS) measured at different positions. The in-gap states with larger spectral weight are situated more closer to the chemical potential, while at the same time the spectral weight of UHB moves to higher energy. This relation is opposite to that found in earlier works \cite{PhysRevB.46.3183,PhysRevLett.77.131,0295-5075-56-4-563,PhysRevLett.80.2393,PhysRevB.73.165114,PhysRevB.81.235133}, where in-gap peak moves to higher energy as doping increases. Finally, the positions are anti-correlated between sites with higher spectral weight for UHB and in-gap states. This is consistent with an effective doping picture. That is, at the position where the UHB has strong intensity, the effective doping is close to zero or no doping and the in-gap states are not seen. On the other hand, at the position where the in-gap states show up, doping is finite and the intensity of the UHB becomes weaker. Note that similar results has been found previously in optical conductivity measurements\cite{uchida1991optical,katsufuji1995spectral,padilla2005constant}, where there are also peaks around the scale of half an eV at low doping which moves to the lower energy as hole concentration increases. The discrepancy between these newly measured weight distribution and its doping dependence with earlier theoretical works has motivated us to examine the theoretical prediction again and more carefully.

In this work we study the spectral evolution of Mott state with hole doping with a variational approach but with explicit presence of the AFM long range order as in the experiment\cite{ye2013visualizing}. We are particularly interested in the spectra of the unoccupied states above the chemical potential. In the strong coupling regime ($U\geq8t$), at half-filling or in the parent compound, each site is already occupied by an electron with spin $1/2$. Hence, when an electron is inserted into the half-filled state, it must create a doubly occupied site (doublon) and with final states in the UHB and there are no  states inside the CT gap. However, after hole doping  when an electron is inserted into the  lattice, there are two possible final states. The original LHB splits into upper and lower spin density wave (SDW) bands by the presence of AFM order. In this case, finite in-gap spectral weight that corresponds to the upper SDW states shows up. These states, according to our calculation, behaves in a similar manner to  the  in-gap states recently found by STS\cite{cai2015visualizing,ye2013visualizing} with respect to the energy scale and evolution.  These in-gap states also have components of states in the UHB with doublons despite the main contribution from upper SDW states. Thus these in-gap states are a mixture of LHB and UHB states  and they absorb most of the spectral weight transferred from UHB. As hole doping increases, in-gap states move toward the chemical potential with increasing spectral weights and the energy separation between UHB and LHB is effectively getting smaller.   
This provides a slightly different version from the ED result\cite{PhysRevB.45.10032} without including the AFM order, which shows that the weights are transferred from UHB to LHB as the holes are doped but the band-edge separation has little dependence on the hole concentration.


Below we first calculate the ground state of a one-band HM in the presence of the AFM order by means of the VMC method. Then several  states  with one electron added to the ground state are proposed. These states contributing to  the unoccupied states or the inverse photoemission spectra (IPES)  are orthogonalized to find the quasi-particle states. Then the spectral weights of these states are all calculated and compared with experiment\cite{cai2015visualizing,ye2013visualizing} with respect to the energy evolution and the spectral weight redistribution. In addition, we also examine our results for different value of U to study the changes from weak to strong coupling.

\section{\label{sec:level1}Formalism and Method}
A well-known model which includes the low energy physics in $CuO_2$ planes is the three-bands HM\cite{PhysRevLett.58.2794}. In this model, the parent compound without any extra doped holes has every Cu in  $3d^{9}$ configuration with a spin $1/2$ hole. This is like a half-filled one-band HM with very large on-site Coulomb repulsion U and every site has a spin $1/2$. When a hole is doped into the $CuO_2$ plane, it resides at the Oxygen site\cite{PhysRevLett.55.418}. Due to the strong super exchange interaction between the Cu spin and the doped hole on Oxygen, Zhang and Rice\cite{PhysRevB.37.3759} found the interaction of two oxygen p orbitals and Cu d orbital leading to three bands, the non-bonding states, anti-bonding triplet states, and the bonding singlet states known as the Zhang-Rice (ZR) singlet. There are large energy differences between the three states and only ZR singlet is assumed to be important for consideration. This ZR singlet in the three-bands model is similar to the vacant site when the hole is doped into the one-band HM. When the energy difference between ZR singlet and the Cu $3d^{10}$ state, which is the effective CT gap,  is not punitively  large, the Cu hole can jump to its neighboring Oxygen to form ZR singlet while the original Cu turns into a $3d^{10}$ configuration without spin. This is similar to the charge fluctuation process in one-band HM to turn the two nearest neighbor opposite spins into the short lived configuration of a doublon-hole pair.   
By making the correspondences of the doublon in one-band HM with the Cu $3d^{10}$, the Hubbard gap with the CT gap and the vacant site or hole with the ZR singlet, we could clarify the physics by studying the one-band HM instead of the more complicated three-bands model\cite{threebandspectrumHorsch}\footnote{Actually, there is still a difference between the two models when we consider charge fluctuation of doublon-hole pairs. In the one-band model there is only one kind of hole hopping processes while there are two kinds in the three-bands model. One process is simply the exchange of ZR singlet with a Cu spin as in the one-band case, the amplitude $t$ was estimated in Ref. \cite{PhysRevB.37.3759}. The other process is the formation of doublon-hole pair or Cu-$3d^{10}$-ZR singlet. This hopping amplitude is about $\bar{t}=0.4t$.}.

The one-band HM we consider is 
\begin{equation}
H=-t\sum\limits_{<i,j>,\sigma}(c^{\dagger}_{i,\sigma}c_{j,\sigma}+h.c.)+\sum\limits_{i}Un_{i,\uparrow}n_{i,\downarrow}
\end{equation}
where t is the hopping integral of a single electron, $<i,j>$ denotes the nearest-neighbour sites.  $U/t$ is usually taken to be 10  if not specified otherwise.

\begin{figure}
\centering
\includegraphics[scale=0.16]{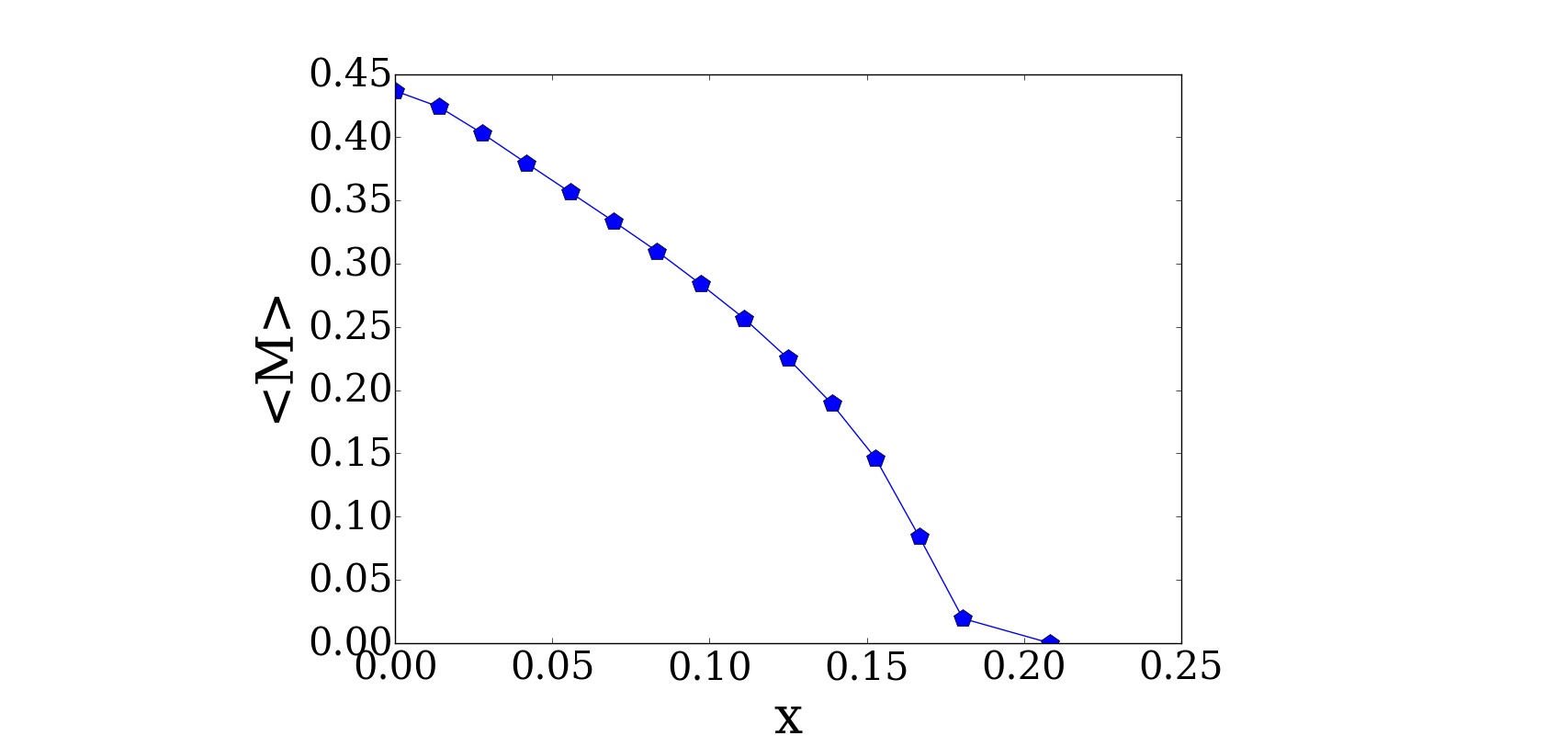}
\caption{The ground state staggered magnetization as a function of hole concentration for U=10t in a 12 by 12 lattice. The AFM order disappears around 0.18 doping in this model.}
\label{fig:groundstate}
\end{figure}

The variational ground  state we chose in the VMC method is the Jastrow type state with coexisting antiferromagnetism and d-wave superconductivity\cite{PhysRevB.55.5983,kobayashi2013coexistence}.
\begin{equation}
\vert\Psi_{variational} \rangle \equiv \hat{P}_{d-h}\hat{P}_{d}\vert\Psi_{afm-ds} \rangle
\end{equation}
where $\hat{P}_{d} = g^{\hat{d}}$ is the Gutzwiller projection operator with $\hat{d} = \sum_{i}\hat{d}_{i}=\sum_{i}\hat{n}_{i\uparrow}\hat{n}_{i\downarrow}$ representing the doublon number. The Gutzwiller factor $g$ suppresses the double occupancy or doublon number when it is less than one. The  Jastrow factor for doublon-hole binding\cite{doi:10.1143/JPSJ.59.3669} is $\hat{P}_{d-h}\equiv\prod_{i}[ 1-Q_{d-h}\hat{d}_{i} \prod_{\tau}(1-\hat{h}_{i+\tau}) ] $
where $\tau$ connects the nearest neighbors and $\hat{h}_{i}\equiv(1-\hat{n}_{i\uparrow})(1-\hat{n}_{i\downarrow})$ is the number of holes on site i. This  factor  ensures
 the insulating phase at half filling. This factor may come in different forms. 
Here we restricted the occurrence of free doublons that aren't bound with holons with a variational parameter $Q_{d-h}\le 1$.   The wave function $\vert\Psi_{afm-ds} \rangle$ with coexisting  antiferromagnetizm and superconductivity has been proposed before\cite{PhysRevB.55.5983},
\begin{align}
\vert\Psi_{afm-ds} \rangle \equiv \hat{P}^{Ne}\prod\limits_{k\in MBZ}&(u_{k-}+v_{k-}\alpha^{\dagger}_{k\uparrow}\alpha^{\dagger}_{-k\downarrow})
\nonumber \\
&(u_{k+}+v_{k+}\beta^{\dagger}_{k\uparrow}\beta^{\dagger}_{-k\downarrow})\vert 0 \rangle
\end{align}
where $\hat{P}^{Ne}$ restricts the state to have $Ne$ electrons. The operators
\begin{align}
&\alpha ^{\dagger}_{k\sigma}\equiv a_{k}c^{\dagger}_{k,\sigma}+\sigma b_{k}c^{\dagger}_{k+Q,\sigma} \nonumber \\
&\beta ^{\dagger}_{k\sigma}\equiv -\sigma b_{k}c^{\dagger}_{k,\sigma}+a_{k}c^{\dagger}_{k+Q,\sigma}
\label{eq:afqp}
\end{align}
correspond to the lower ($\alpha$) and upper ($\beta$) spin density wave states (SDW) with coefficients $a_{k}^2 \equiv \frac{1}{2}(1-\frac{\varepsilon_{k}}{\sqrt{\varepsilon_{k}^2+M_{v}^2}}),b_{k}^2 \equiv 1-a_{k}^2$, $M_{v}$ being a variational parameter proportional to staggered magnetization. Here we consider commensurate SDW and Q is chosen to be $(\pi, \pi)$, and  $k$ is within the magnetic Brillouin zone (MBZ). The coherent coefficients $u_{k\pm}$ and $v_{k\pm}$ are defined by $u_{k\pm}^2\equiv \frac{1}{2}(1-\frac{(E_{k\pm}-\mu)}{\sqrt{\Delta_{k}^2+(E_{k\pm}-\mu)^2}})$ and $v_{k\pm}^2\equiv 1-u_{k\pm}^2$, respectively. 
The plus/minus sign denotes upper/lower SDW states, $E_{k\pm}\equiv \pm \sqrt{M_{v}^2+\varepsilon_{k}^2}$ is the mean field SDW energy with $\varepsilon_k\equiv-2t(\cos{k_x}+\cos{k_y})$. The chemical potential $\mu$ is also taken to be a variational parameter. Finally, $\Delta_k=\Delta(\cos{k_x}-\cos{k_y})$ is the d-wave gap. For numerical convenience, our boundary condition is chosen to be periodic in x direction and anti-periodic in y direction. The staggered magnetization obtained for the ground states is plotted as a function of hole concentration in Fig.~\ref{fig:groundstate}. The result that AFM order disappears around $18\%$ agrees with Ref. \cite{kobayashi2013coexistence} even though they have used a more sophisticated trial wave function\footnote{This result overestimates the doping range for the AFM phase. To be consistent with cuprate phase diagram, we need much larger U/t. Since we are only interested in the AFM state, this discrepancy does not change our main conclusion.}.
By adding an electron to the ground state we can now calculate the IPES. We shall consider the simplest quasi-particle states and there are four kinds for each k point within the MBZ.
\begin{align}
\vert 1_{k,\sigma} \rangle \equiv &\sum\limits_{i}\hat{n}_{i,\bar{\sigma}}(e^{ikR_{i}}c^{\dagger}_{i,\sigma})\vert g_{N} \rangle \label{eq:qp1}\\
\vert 2_{k,\sigma} \rangle \equiv &\sum\limits_{i}\hat{n}_{i,\bar{\sigma}}(e^{i(k+Q)R_{i}}c^{\dagger}_{i,\sigma})\vert g_{N} \rangle \\
\vert 3_{k,\sigma} \rangle \equiv &\sum\limits_{i}\lbrace(1-\hat{n}_{i,\bar{\sigma}})\prod\limits_{\tau}[1-\hat{d}_{i+\tau}\prod\limits_{\tau’\neq -\tau}(1-\hat{h}_{i+\tau+\tau’})]\nonumber \\&(e^{ikR_{i}}c^{\dagger}_{i,\sigma})\rbrace\vert g_{N} \rangle \\
\vert 4_{k,\sigma} \rangle \equiv &\sum\limits_{i}\lbrace(1-\hat{n}_{i,\bar{\sigma}})\prod\limits_{\tau}[1-\hat{d}_{i+\tau}\prod\limits_{\tau’\neq -\tau}(1-\hat{h}_{i+\tau+\tau’})]\nonumber \\&(e^{i(k+Q)R_{i}}c^{\dagger}_{i,\sigma})\rbrace\vert g_{N} \rangle \label{eq:qp4}
\end{align}
where $\vert g_{N} \rangle$ denotes the ground state with particle number N. States $\vert 1 \rangle$ and $\vert 2 \rangle$ both create an extra doublon in the ground state, so they belong to the UHB in the atomic limit. On the contrary, states $\vert 3 \rangle$ and $\vert 4 \rangle$ add an electron to a vacant site, and they belong to the LHB. Note that at low hole concentration, there are finite doublon-hole bound pairs generated by quantum fluctuation. If we add an electron to the hole site bound with a doublon, it would create a free doublon.  these states are also in the UHB, which we had confirmed by direct calculation of their energy. These states have large overlaps with $\vert 1 \rangle$ and $\vert 2 \rangle$ and they also contribute very little spectral weight which is proportional to the doublon number.  So without loss of generality we shall exclude the process of creating free doublon from states $\vert 3 \rangle$ and $\vert 4 \rangle$ .

To find the eigenstate within the chosen basis, for each k point in MBZ, we calculated the Hamiltonian matrix element by Monte Carlo algorithm $\langle H(k,\sigma) \rangle_{ij}=\langle i_{k,\sigma} \vert H \vert j_{k,\sigma} \rangle, (i,j=1\sim4)$. Since it is a non-orthonormal basis, we also need the metric tensor, $\langle G(k,\sigma) \rangle _{ij}=\langle i_{k,\sigma} \vert j_{k,\sigma} \rangle$. Next, we solve the 4 by 4 generalized eigenvalue problem and obtain four eigenstates with an extra quasi-particle  $\vert \Psi^{i} _{N+1}(k,\sigma) \rangle, (i=1\sim 4)$ and their energy $E^{i}_{N+1}(k,\sigma)$. The energy to insert a quasi-particle is defined as 
\begin{equation}
\xi^{i+}(k,\sigma)=E^{i}_{N+1}(k,\sigma)-E_{N+1,min}
\end{equation}
where the minimum eigenenergy $E_{N+1,min}$ is considered to be at the chemical potential. 

For each eigenstate at k, the spectral weight of inserting a particle contains two contributions $Z^{+}(k,\sigma)$ and $Z^{+}_{Q}(k,\sigma)$, which are defined by
\begin{align}
&Z^{i+}(k,\sigma)=\vert \langle {\Psi}^{i}_{N+1}(k,\sigma)\vert c^{\dagger}_{k,\sigma}\vert{g}_{N}\rangle\vert^2 \\
&Z^{i+}_{Q}(k,\sigma)=\vert \langle {\Psi}^{i}_{N+1}(k,\sigma)\vert c^{\dagger}_{k+Q,\sigma}\vert{g}_{N}\rangle\vert^2
\end{align}.

A similar procedure is also applied to study states with an electron removed  from the ground state, this is for the photoemission spectra (PES). The states can be simply obtained by a transformation $c^{\dagger} \rightarrow c$. The basis states are
\begin{align}
\vert 1^-_{k,\sigma} \rangle \equiv &\sum\limits_{i}\hat{n}_{i,\sigma}(e^{-ikR_{i}}c_{i,\bar{\sigma}})\vert g_{N} \rangle \\
\vert 2^-_{k,\sigma} \rangle \equiv &\sum\limits_{i}\hat{n}_{i,\sigma}(e^{i(-k+Q)R_{i}}c_{i,\bar{\sigma}})\vert g_{N} \rangle \\
\vert 3^-_{k,\sigma} \rangle \equiv &\sum\limits_{i}\lbrace(1-\hat{n}_{i,\sigma})\prod\limits_{\tau}[1-\hat{d}_{i+\tau}\prod\limits_{\tau’\neq -\tau}(1-\hat{h}_{i+\tau+\tau’})]\nonumber \\&(e^{-ikR_{i}}c_{i,\bar{\sigma}})\rbrace\vert g_{N} \rangle \\
\vert 4^-_{k,\sigma} \rangle \equiv &\sum\limits_{i}\lbrace(1-\hat{n}_{i,\sigma})\prod\limits_{\tau}[1-\hat{d}_{i+\tau}\prod\limits_{\tau’\neq -\tau}(1-\hat{h}_{i+\tau+\tau’})]\nonumber \\&(e^{i(k+Q)R_{i}}c_{i,\bar{\sigma}})\rbrace\vert g_{N} \rangle.
\end{align}
After diagonalization we have four eigenstates with a particle removed  $\vert \Psi^{i-} _{N-1}(k,\sigma) \rangle, (i=1\sim 4)$ from the ground state and their corresponding energies are $E^{i-}_{N-1}(k,\sigma)$ for each k. The energy to remove a particle  becomes
\begin{equation}
\xi^{i-}(k,\sigma)=-E^{i}_{N-1}(k,\sigma)+E_{N-1,min}
\end{equation}
Similarly, the minimum eigenenergy $E_{N-1,min}$  is considered to be at the chemical potential. The spectral weights to remove a particle  are related to $Z^{-}(k,\sigma)$ and $Z^{-}_Q(k,\sigma)$ defined by
\begin{align}
&Z^{i-}(k,\sigma)=\vert \langle {\Psi}^{i}_{N-1}(k,\sigma)\vert c_{-k,\sigma}\vert{g}_{N}\rangle\vert^2 \\
&Z^{i-}_{Q}(k,\sigma)=\vert \langle {\Psi}^{i}_{N-1}(k,\sigma)\vert c_{-k+Q,\sigma}\vert{g}_{N}\rangle\vert^2
\end{align}

Finally we can combine the PES and IPES together,
\begin{align}
\rho(\omega)\equiv \frac{1}{d\omega}&\sum\limits_{\omega < \xi^{i\pm}(k,\sigma) < \omega + d\omega}(Z^{i+}(k,\sigma)
\nonumber \\
&+Z^{i+}_{Q}(k,\sigma)+Z^{i-}(k,\sigma)+Z^{i-}_{Q}(k,\sigma))
\end{align}
The summation is over $i,k$ and $\sigma$. All results reported here are carried out  on a 12 by 12 lattice.

\section{\label{sec:level1}Results}
\begin{figure}[t]
  \centering
    \includegraphics[width=0.5\textwidth]{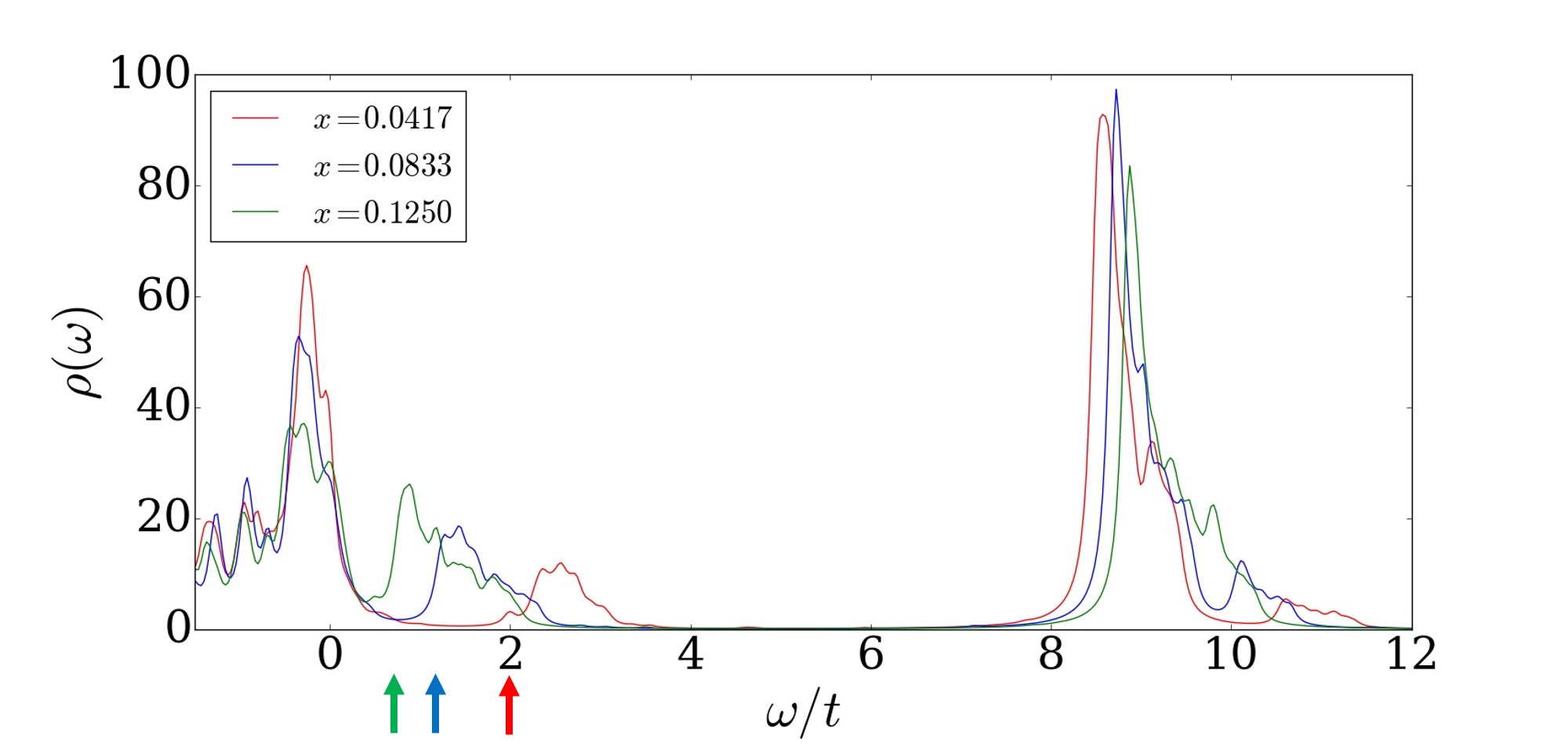}
\caption{Spectral function for doping $x=0.0417, 0.0833$, and $0.125$. A Lorentzian broadening with width $\Gamma=0.15t$ is applied to the delta functions for the eigenstates. The arrows indicate the value of $\xi_{0}$ as defined in the text.}
\label{fig:spectrum_new}
\end{figure}

The spectral function for $U/t=10$ is plotted as a function of energy for three hole concentrations in Fig.~\ref{fig:spectrum_new}. A sharp UHB peak around $9t$ is seen. More importantly, a broad band of in-gap states appears in the range $0.1U \sim 0.35U$ which we believed to be mostly unoccupied upper SDW states. These states are about the same energy range as the newly found signals in the STS\cite{cai2015visualizing,ye2013visualizing}. As hole concentration increases, the UHB weight shifts toward higher energy while the in-gap states moves toward the chemical potential, {\it i.e.} the energies of in-gap states decrease. The spectral weights of these in-gap states also increase with doping. There is clearly a spectral weight transferred from  UHB to low energy states (LES) that are between chemical potential and UHB. To verify the relationship between in-gap states and the upper SDW states, we calculate the inner product between these states. The upper SDW states can be constructed in the same way as wave functions $\vert 3 \rangle$ and $\vert 4 \rangle$ except now we restrict the electron to be inserted into the upper SDW band defined in Eq. (\ref{eq:afqp}),
 
\begin{align}
&\vert \psi_{u-SDW}(k,\sigma)\rangle\equiv\sum\limits_{i}\lbrace(1-\hat{n}_{i,\bar{\sigma}}) \nonumber \\
&\prod\limits_{\tau}[1-\hat{d}_{i+\tau}\prod\limits_{\tau’\neq -\tau}(1-\hat{h}_{i+\tau+\tau’})](\langle i \vert \beta_{k}\rangle_{\sigma} c^{\dagger}_{i,\sigma})\rbrace\vert g_{N} \rangle
\end{align}
\begin{figure}
  \centering
    \includegraphics[width=0.5\textwidth]{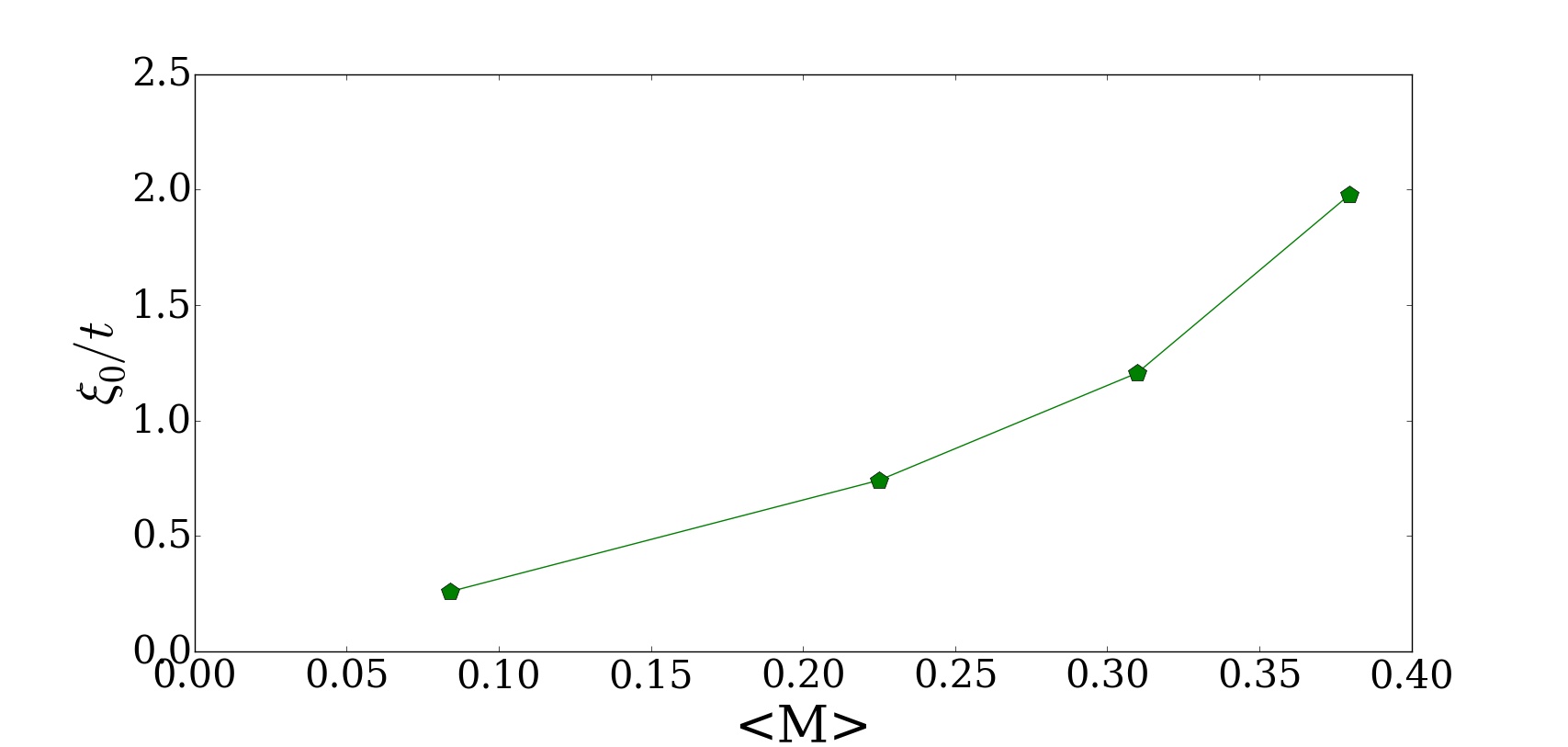}
\caption{$\xi_0$ of the upper SDW states as a function of magnetization. The energy together with magnetization are collected from four different doping $x=0.0417, 0.0833, 0.125,$ and $0.1667$, here magnetization is inversely proportional to hole concentration.}
\label{fig:u-SDWE_edge}
\end{figure}

Where $\langle i \vert \beta_{k}\rangle_{\sigma}=-\sigma b_{k}e^{ikR_{i}}+a_{k}e^{i(k+Q)R_{i}}$. At each of the k points, we found large overlap ($>0.8$) between $\vert \psi_{u-SDW}(k,\sigma)\rangle$ and the in-gap states. Thus these in-gap states are essentially the upper SDW states although there are contributions from states $\vert 1_{k,\sigma}\rangle$ and $\vert 2_{k,\sigma}\rangle$ which are in the UHB. Besides these states near the chemical potential, there are also contributions from lower SDW states that are now vacant due to hole doping. The energy scale of these states is roughly determined by the coupling between the states $\vert 3_{k,\sigma}\rangle$ and $\vert 4_{k,\sigma}\rangle$. Considering the transition between upper and lower SDW states, our result gives a possible explanation of the half an eV peak in optical conductivity measurements\cite{uchida1991optical,katsufuji1995spectral,padilla2005constant} that shows a decreasing absorption  energy and an increasing weight  with more doping. This will be left for future works.

In the presence of AFM long range order, staggered magnetization opens a gap between the upper and lower SDW states.   In our case, the effective staggered magnetization is proportional to the variational magnetic field $M_{v}$. To illustrate this relation, we define $\xi_0$ by the lowest eigenenergy of the quasi-particle states that has an inner product with $\vert \psi_{u-SDW}(k,\sigma)\rangle$ larger than 0.8\footnote{Actually, the value of $\xi_0$ remains the same even if we change this criterion to 0.7 or 0.9}, which provides a good indicator of the lower edge of the in-gap states. Thus, positive correlation is expected between the upper SDW band edge $\xi_0$ and AFM strength $\langle M\rangle$ as shown in Fig.~\ref{fig:u-SDWE_edge}. 
Since $\langle M\rangle$ is inversely correlated with doping, this gives a natural explanation of the  reduction of energies of in-gap states as doping increases.  
\begin{figure}
  \centering
    \includegraphics[width=0.5\textwidth]{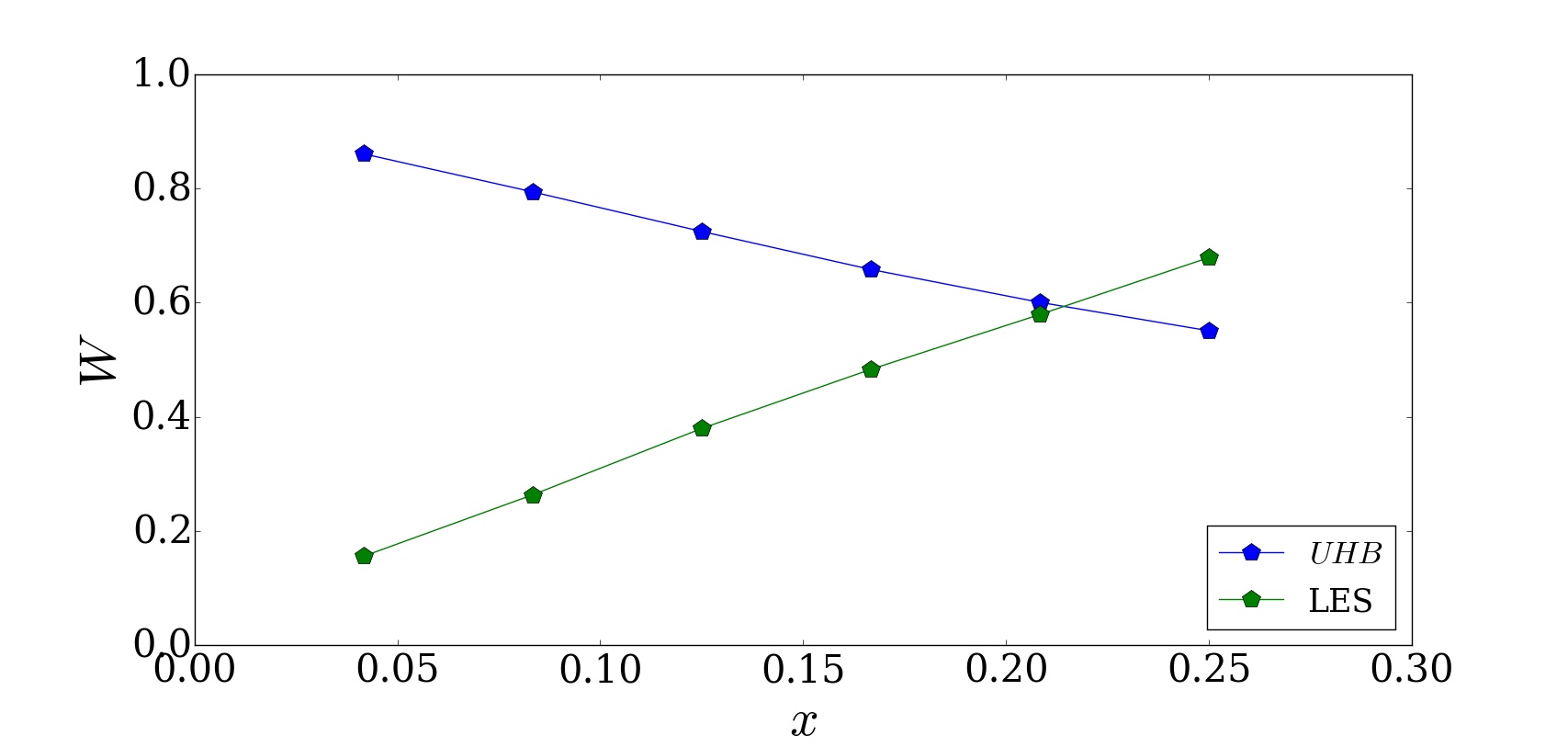}
\caption{Total spectral weight for UHB and LES. In under-doped regime, $W_{UHB}$ and $W_{LES}$ evolve linearly. By fitting, we found the slope to be $-1.66$ for $W_{UHB}$ and $2.69$ for $W_{LES}$.}
\label{fig:SpecTrans}
\end{figure}
Next we shall examine the spectral weight transferred from UHB to LES. We calculated the total weight $W_{UHB}$ and $W_{LES}$. 
\begin{align}
W_{UHB} \equiv \sum \limits_{\xi^{i+}(k,\sigma)>4t}Z^{i+}(k,\sigma)+Z^{i+}_Q(k,\sigma) \nonumber \\
W_{LES} \equiv \sum \limits_{\xi^{i+}(k,\sigma)\leq 4t}Z^{i+}(k,\sigma)+Z^{i+}_Q(k,\sigma)
\end{align}
$W_{UHB}$ and $W_{LES}$ at different doping are shown in Fig.~\ref{fig:SpecTrans}. This agrees quantitatively with the previous results from ED\cite{PhysRevB.45.10032}, ED + cluster DMFT \cite{PhysRevB.80.165126}, and CPT\cite{PhysRevLett.108.076401} despite the fact that AFM order was not considered in these works. This shows that the appearance of AFM order doesn't affect total spectral weight transferred as the sum rule should be satisfied. 

If we consider each site at the atomic limit, UHB is completely unoccupied at half-filling, hence it has a spectral weight equal to 1. This weight $W_{UHB}$ reduces to $1-x$ when $x$ holes are doped into the system. Upon doping, an electron could be added to the empty or hole site in two choices from the spin degrees of freedom, hence the spectral weight for $W_{LES}$ is $2x$. However, it is known that beyond this atomic limit there should be a dynamical correction\cite{PhysRev.157.295} that comes from the coupling between these states which enhances the weight transfer and would give $W_{UHB}=1-x-\alpha$ and $W_{LES}=2x+\alpha$. According to our calculation, the renormalization $\alpha$ at $U/t=10$ is around $0.67x$.
\begin{figure}
\centering
\subfigure{
\includegraphics[scale=0.13]{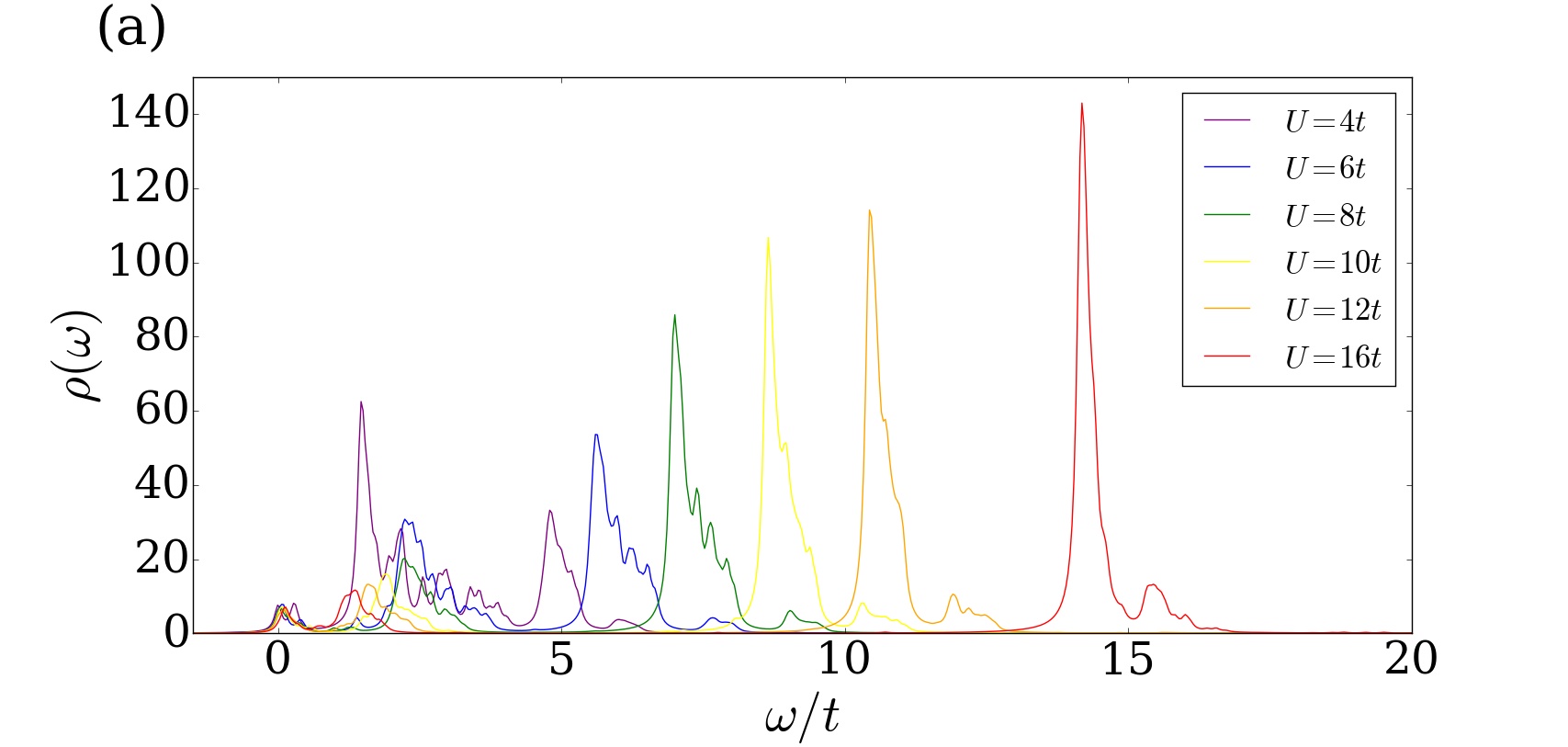}}
\subfigure{
\includegraphics[scale=0.13]{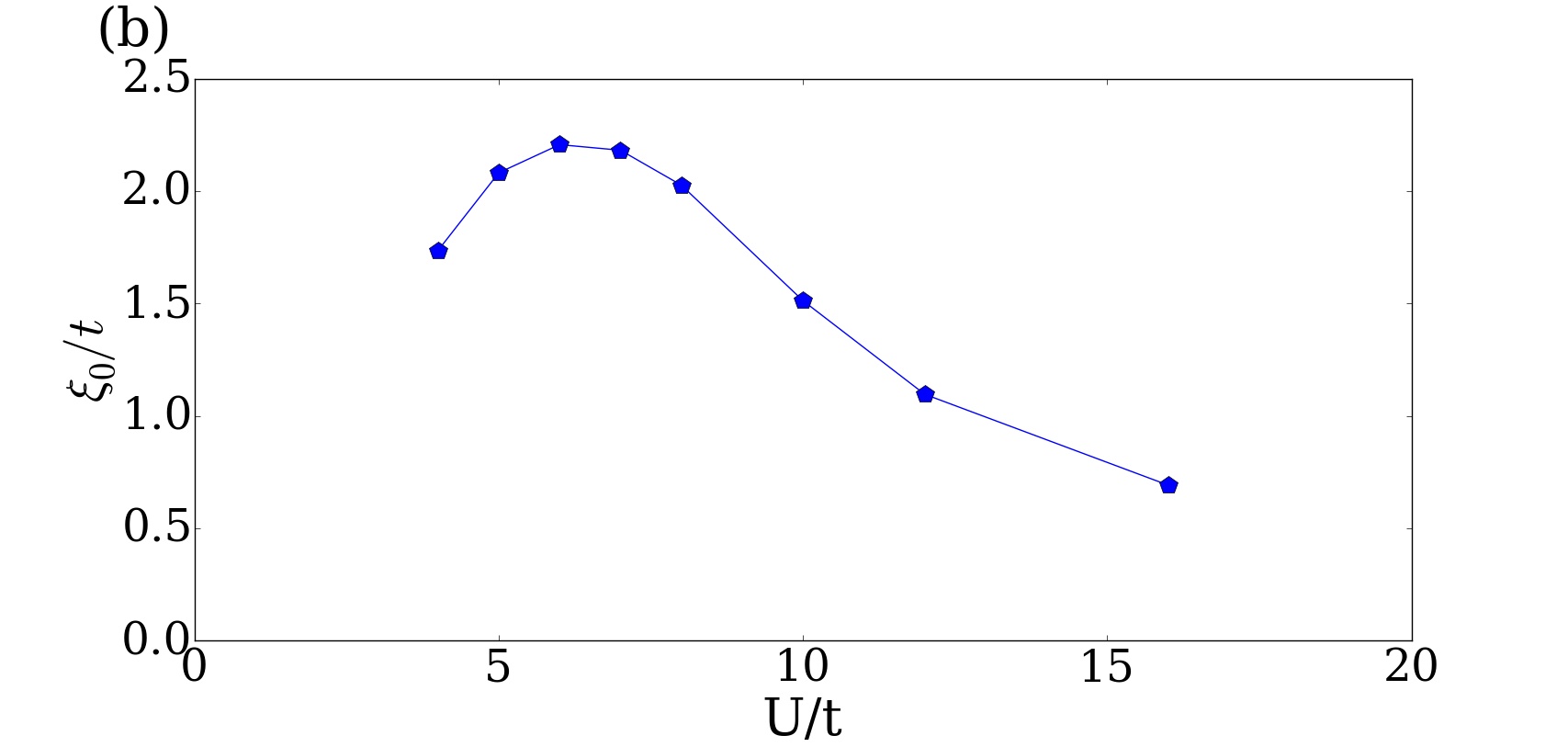}}
\subfigure{
\includegraphics[scale=0.13]{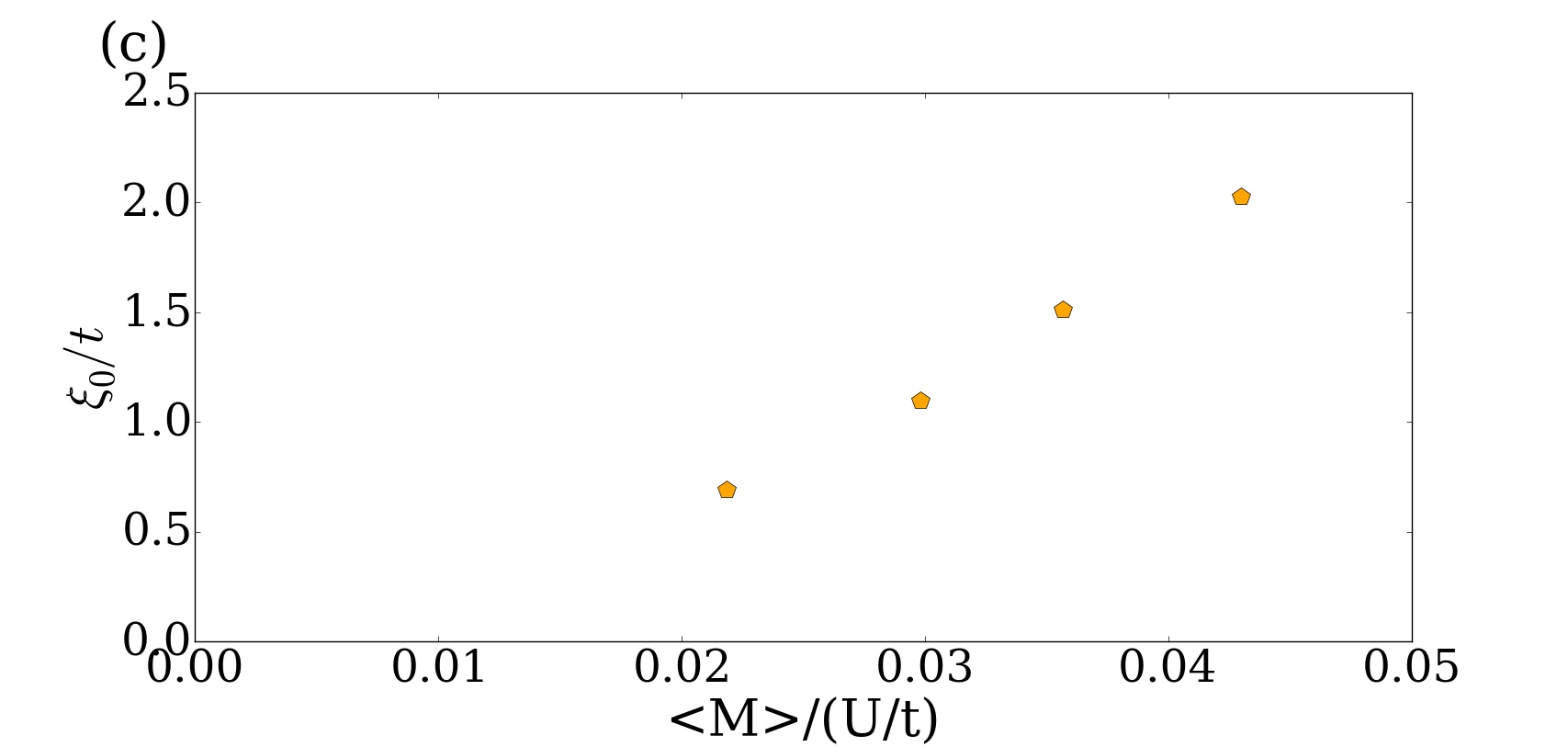}}
\caption{(a)IPES for different U at $x=0.0556$. (b)$\xi_{0}/t$ at different U. (c)$\xi_{0}/t$ as a function of $\langle M \rangle /(U/t)$ for $U\geq 8t$. This shows that the virtual exchange interaction $J$ accounts for the results.}
\label{fig:differentU} 
\end{figure}

Since the AFM order depends on U, it is important to examine the evolution of spectral functions with different values of U. In Fig.~\ref{fig:differentU}(a), the spectral functions for inserting a quasi-particle in the ground state are plotted for a range of U values. For $U/t\geq 6$, the separation between UHB and LES is clear and as expected, UHB energy scales with U. The trend suggests that as U becomes smaller, weight of in-gap state at $x=0.0556$ becomes larger and their component of $\vert 1\rangle$ and $\vert 2 \rangle$ increases. For weak or intermediate $U/t$ the weight of the in-gap state is comparable or even larger than that of UHB.   To further examine the U dependence of these spectra, we plot the upper SDW band edge $\xi_0$ as a function of U in Fig.~\ref{fig:differentU}(b).  Before U reaches $U_c \sim 8t$ to enter the Mott region, the $\xi_0$ is proportional to $U/t$. This is expected from a mean-field treatment for a  weak or intermediate coupling U in the one-band HM, as the gap due to AFM order is proportional to $\langle M\rangle U$. Once the system enters the Mott region with U greater than $U_c$, it is the superexchange interaction $J$ that determines the AFM order. Hence  $\xi_0 \sim \langle M\rangle J \sim \langle M\rangle /(U/t)$. $\xi_0$ plotted as a function of  $\langle M\rangle /(U/t)$ in Fig.~\ref{fig:differentU}(c) for  $U\geq U_c$.  This shows that one could expect a maximum energy for in-gap states when U/t is near the critical value for a Mott transition. 
\begin{figure}
\centering
\subfigure{
\includegraphics[scale=0.14]{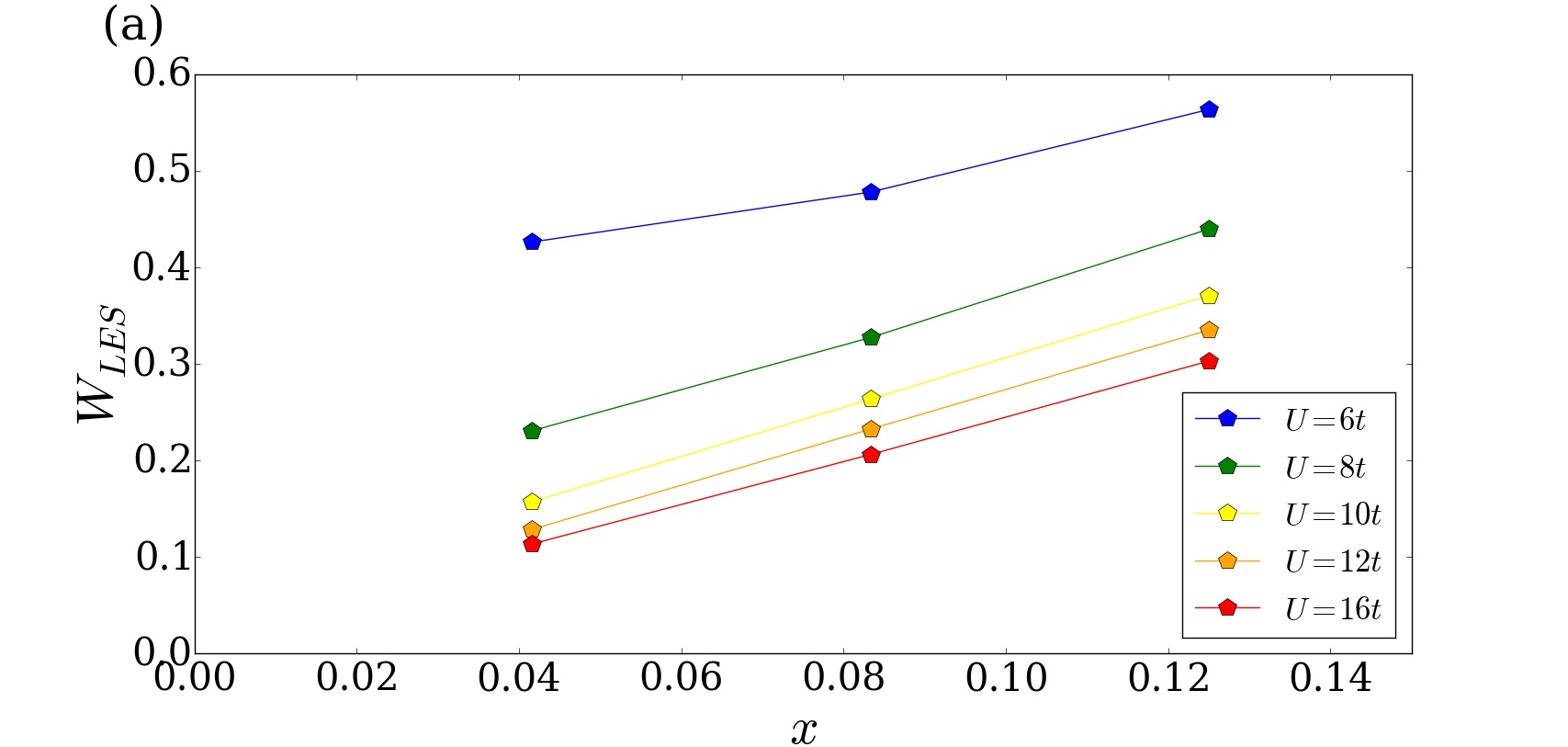}}
\subfigure{
\includegraphics[scale=0.14]{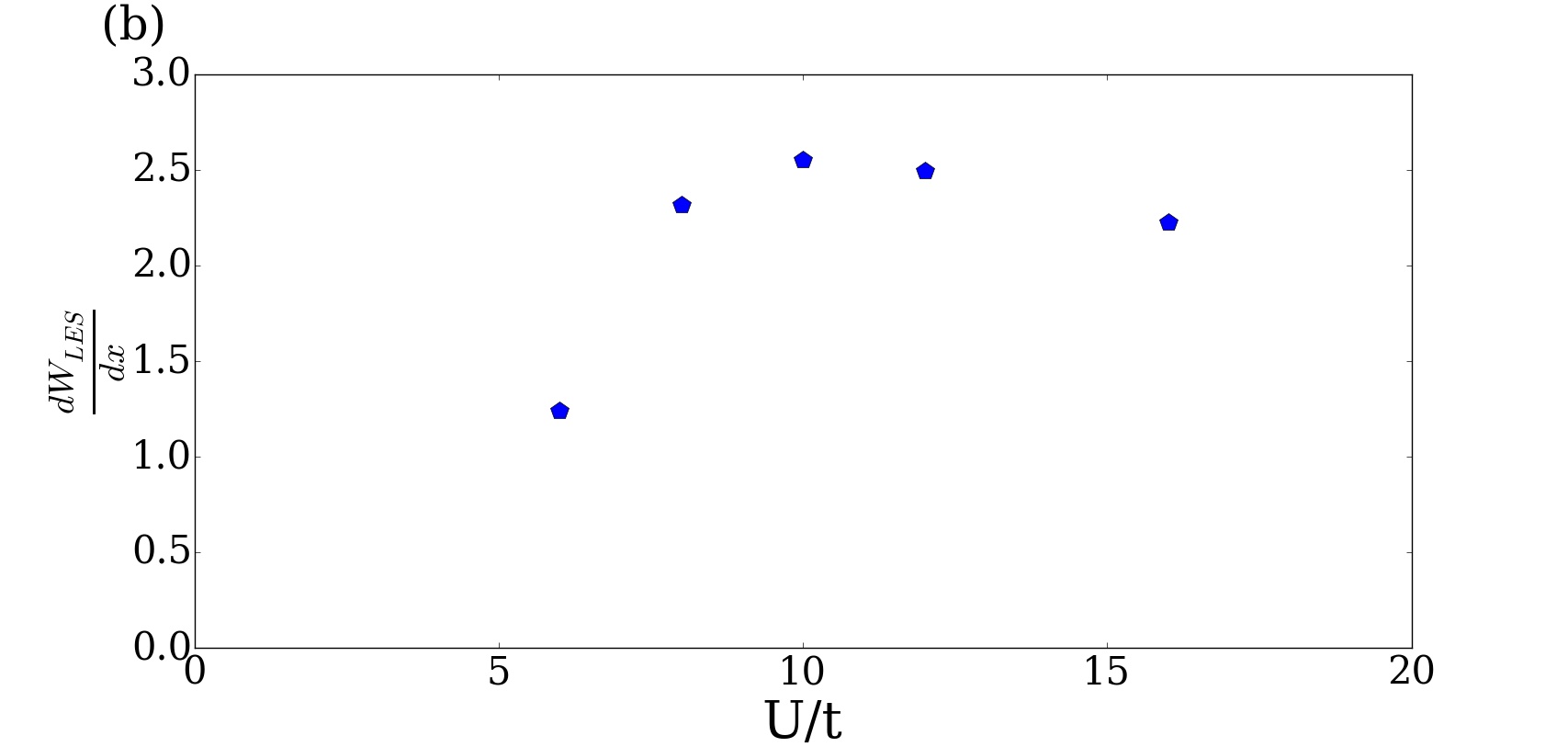}}
\caption{(a)$W_{LES}$ for different U at $x=0.0417$, $0.0833$, and $0.125$. (b)Slope of $W_{LES}$ for different U taken from data at $x=0.0417$ and $0.0833$.}
\label{fig:SpecTrans_DiffU} 
\end{figure}

Now let’s examine the spectral weight for LES.  As shown in Fig. \ref{fig:SpecTrans_DiffU}(a), $W_{LES}$ is extrapolated to zero at half-filling for U much larger than $U_c$. It has a sudden increase below $U/t=6$.  This suggests that these in-gap weight won't disappear at half-filling in intermediate coupling case. Moreover, as U decreases to the weak coupling regime, as shown in the $U/t=4$ case in Fig. ~\ref{fig:differentU}(a),   the major peak appears at the location of upper SDW states. Those upper SDW states begin to merge with the UHB and the excitation gap becomes an SDW gap and there is no clear separation between UHB and upper SDW states. This is in accordance with the mean field theory.  Therefore, the disappearance of SDW states at half-filling as U increases is a characteristic of Mott transition. Furthermore we can also examine the slope of the $W_{LES}$.  An important feature of Mott insulator is the spectral weight of  $2x+\alpha$ for the LES. In the infinite U limit, according to Ref. \cite{PhysRev.157.295}, $\alpha$ is proportional to $1/U$. In this regime, as U decreases, there is an increase in the slope of $W_{LES}$. In the free particle limit, on the other hand, there is no UHB and all of the weight $1+x$ is distributed close to chemical potential. The transition between strong and weak coupling can be clearly seen in Fig.~\ref{fig:SpecTrans_DiffU}(b)., starting from $U/t=16$, as U decreases, there is an increase in slope of $W_{LES}$. As soon as U crosses the Mott transition ($U_c$), the slope has a sudden drop. At $U/t=6$, the slope is much less than $2$, suggesting that it has crossed $U_c$ and Mott physics no longer applies.



\section{\label{sec:level1}Conclusion}
In summary, by using a variational approach to study the HM in the AFM phase, we construct several quasi-particle states to study the evolution of PES and, in particular, IPES, with hole doping. The substantial amount of spectral weight inside the Mott gap is due to the mostly unoccupied upper SDW bands. These states interact strongly with the UHB so that there is a large spectral weigh transferred from UHB to these in-gap states. Although we use one-band HM instead of specific oxygen orbitals, we are able to capture the detail evolution of the spectrum with doping observed in the recent STS on cuprates.   Our results also agree  with  previous numerical works  with respect to the dynamical spectral-weight transfer with a renormalization of about $\alpha \sim 0.65x$  at $U/t=10$. This agreement is a bit surprise as previous works have not included AFM order\cite{PhysRevB.45.10032,PhysRevB.80.165126}. Maybe this is due to the fact that for IPES, there is a much smaller incoherent spectral weight\cite{PhysRevLett.95.137001} and quasi-particle states we considered have almost all the spectral weights.

We like to point out that in the lightly doped cuprate samples\cite{cai2015visualizing} in experiments,  there is no density of states at the chemical potential.  Since the sample is quite inhomogeneous, we believe that there is a strong localization effect that depletes the density of states at chemical potential.  Since we have not considered the localization or disorder  in our calculation, our result has  a small peak near chemical potential when an electron is added to the lower-SDW states. Fortunately, the in-gap states that we focused on has energy much larger than this localization gap, so that we can account for them without considering the disorder. An interesting possibility of localization may be due to the checkerboard pattern observed in\cite{cai2015visualizing}. In the strong coupling limit of the HM or the $t-J$ model, it is recently shown\cite{tu2016genesis} that at very low doping, there are states with checkerboard patterns involving SDW.  Hence our account of SDW as the source of in-gap states may be a reasonable approximation. Further work on this is underway.

Our result also shows that the spectral distribution has a non-monotonic behavior when  U is increased above $U_c$ to enter the Mott physics. At half-filling, there is a small SDW gap at small U, and also a finite spectral weight at low energy ($W_{LES}$) above chemical potential. But the gap  changes to the much larger Hubbard gap as U becomes larger than  $U_c$ and there is absolutely no spectral weight within the gap. After doping the energy of in-gap states changes from mean field type ($\propto U$) to t-J ($\propto 1/U$) type and left a peak around $U_c$. At small doping the low energy spectral weight above chemical potential is proportional to doping for $U>U_c$. These properties might be useful  to observe the Mott transition by applying pressure.

\bibliography{draft-rev-7}
\end{document}